\documentclass[singlespacing]{elsart}

 \usepackage{graphicx}

\usepackage{amssymb}

\begin{document}

\begin{frontmatter}



 \title{Entanglement in Weisskopf-Wigner theory of atomic decay in free space}


 \author{J. F. Leandro},
 \author{F. L. Semi\~ao},
 \address{Departamento de F\'isica, Universidade Estadual de Ponta Grossa - Campus Uvaranas, 84030-900 Ponta Grossa, PR, Brasil}

\begin{abstract}
\textbf{In this paper, we use the Weisskopf-Wigner theory to study the entanglement in the state of the free-space radiation field produced from vacuum due to atomic decay. We show how bipartite entanglement is shared between different partitions of the radiation modes. We investigate the role played by the size of the partitions and their detuning with the decaying atom. The dynamics of the atom-field entanglement during the atomic decay is also briefly discussed. From this dynamics, we assert that such entanglement is the physical quantity that fix the statistical atomic decay time.}
\end{abstract}

\end{frontmatter}


Entanglement plays a central role in quantum information science where it is seen as an important physical resource for information processing beyond the achievable with classical correlations \cite{dense,tele,martin}. Investigations in quantum phase transitions \cite{qpt,qpt2} and statistical mechanics have been realized from the point of view of entanglement \cite{popescu}. In this approach, concepts like randomness, ensemble-averaging or time-averaging are not required. Instead, thermalization results from entanglement between system and environment. Connections between matter and quantum information theory have also been discussed \cite{lloyd,eisert}. Such studies have pointed out that entanglement seems to be important in other areas of physics besides pure quantum information. It is exactly in this scope that this work is introduced.

Considering entanglement as a legitimate physical quantity, this paper is intended to study a fundamental process namely the atomic spontaneous emission. It lies at the core of matter-radiation interaction. The successful description of atomic decay in free space is one of the remarkable achievements of the quantum theory of radiation \cite{ww,scully,petrosyan}. Why does an excited atom decay? It is clear that an isolated atom would never decay from one excited state to another with lower energy because both are eigenstates of the system Hamiltonian. There must be some physical system to couple to the atom in order drive the electronic transition. This external agent is the free space electromagnetic field whose zero-point energy fluctuations are able to cause the atom to decay. In the language of quantum information theory, the entanglement between atom and field is then the responsible for the atomic decay. In what follows, entanglement is studied in the spontaneous emission phenomenon.

We are particularly interested in the entanglement properties of quantum fields. In this paper, the quantized field is the bosonic free-space continuous electromagnetic field. We will analyze the entanglement properties of this field after atomic decay. Entanglement in discretized bosonic fields have already been studied. In particular, the entanglement properties of the ground state and thermal states of this system is studied in detail in \cite{harmonic}. Discrete versions of real free Klein-Gordon fields have also been studied from the point of view of entanglement \cite{kg}. In this study, the relation between entanglement, entropy, and area for a specific harmonic lattice  whose  continuum limit lead to the Hamiltonian of the real Klein-Gordon field is analyzed in great detail. These papers triggered many others studies that also investigated entanglement in discretized quantum fields \cite{wolf,unanyan,gioev,barthel}.

The starting point of the present work is the Weisskopf-Wigner theory of spontaneous emission which is now briefly presented \cite{ww,scully,petrosyan}. In the rotating wave approximation, a two-level atom interacts with the free-space electromagnetic field according to the interaction picture Hamiltonian \cite{scully}
\begin{eqnarray}\label{H}
\hat{H}=\hbar\sum_{\bf{k}}[g_{{\bf{k}}}^{*}({\bf{r}}_0)\sigma_+\hat{a}_{{\bf{k}}}e^{i(\omega-\nu_k)t}+{\rm{H.c.}}],
\end{eqnarray}
where $\omega$ is the angular frequency of the atomic transition (excited state $|a\rangle$ and ground state $|b\rangle$), ${\bf{r}}_0$ is the position of the atom, $\hat{a}_{\bf{k}}$ is the annihilation operator for the field mode $\{{\bf{k}}\}$ (angular frequency $\nu_k$), and
\begin{eqnarray}
g_{\bf{k}}({\bf{r}}_0)=g_{\bf{k}}e^{-i{\bf{k}}\cdot{\bf{r}}_0},
\end{eqnarray}
where
\begin{eqnarray}
g_{\bf{k}}=-\sqrt{\frac{\hbar\nu_{k}}{2\epsilon_0V}}\frac{\bf{\wp}_{ab}\cdot\hat{\epsilon}_{\bf{k}}}{\hbar},
\end{eqnarray} 
with $V$ a quantization volume, $\epsilon_0$ the electric permittivity of free space, $\wp_{ab}$ the dipole moment for the atomic transition, and $\hat{\epsilon}_{\bf{k}}$ the polarization vector of the mode $\{{\bf{k}}\}$. It will be assumed that initially the atom is in the excited state $|a\rangle$ and the field modes are in the vacuum $|0\rangle=|0,0,...\rangle$. According to (\ref{H}) the system evolved state will be
\begin{eqnarray}\label{psit}
|\psi(t)\rangle=c_a(t)|a,0\rangle+\sum_{{\bf{k}}}c_{b,{\bf{k}}}(t)|b,1_{{\bf{k}}},\{0\}\rangle,
\end{eqnarray}
where $|1_{{\bf{k}}},\{0\}\rangle$ represents the field state with one photon in the mode $\{{\bf{k}}\}$ and the rest in the vacuum, and
\begin{eqnarray}\label{coef}
c_a(t)&=&e^{-\Gamma t/2},\\
c_{b,{\bf{k}}}(t)&=& g_{\bf{k}}({\bf{r}}_0)\frac{1-e^{i(\omega-\nu_k)t-\Gamma t/2}}{(\nu_k-\omega)+i\Gamma/2},
\end{eqnarray}
with $\Gamma$ being the free-space atomic decay constant which is giving by 
\begin{eqnarray}\label{ga}
\Gamma=\frac{1}{4\pi\epsilon_0}\frac{4\omega^3\wp_{ab}^2}{3\hbar c^3}.
\end{eqnarray}
In order to obtain the above equations, it was considered that the intensity of the light associated with the emitted radiation is very centered about the atomic frequency $\omega$. This is the essence of the Weisskopf-Wigner theory. In this theory, the free space modes act as an immediate response reservoir, i.e., the atomic spontaneous emission is seen as a Markovian process. 

Now, the entanglement content in the field state after spontaneous decay of the atom is studied in detail. This state is denoted $|\gamma_0\rangle$ and it is obtained from (\ref{psit}) by assuming $t\gg \Gamma^{-1}$
\begin{eqnarray}\label{g0}
|\gamma_0\rangle=\sum_{{\bf{k}}}\frac{g_{\bf{k}}({\bf{r}}_0)}{(\nu_k-\omega)+i\Gamma/2}|1_{{\bf{k}}},\{0\}\rangle.
\end{eqnarray}
It is worth noticing that the state $|\gamma_0\rangle$ is, from the point of view of quantum information science, a member of an important class of multipartite entangled states called generalized $W$ states \cite{w}. However, we must take care when using this state. Although state (\ref{g0}) is presented as a discrete summation over $\bf{k}$, any kind of calculation using it is to be done transforming it to an integral, i.e. an continuum of modes.

The state (\ref{g0}) represents all modes of the free space radiation field, and it is a superposition of the different possibilities of distributing one photon (emitted by the atom) between the infinity of modes. Consequently, this is an entangled multipartite state whose bipartite entanglement between different partitions of radiation modes is now going to be investigated. There are many ways of partitioning the free space modes in two partitions. We think it is physically appealing to choose one partition formed by a central mode with frequency $\nu_{\bf{q}}$ and modes distributed in the interval $(\nu_{\bf{q}}-\epsilon,\nu_{\bf{q}}+\epsilon)$ (let us call it partition $A$), and the other partition formed by the rest (partition $B$). This is an interesting physical choice since it allows us to check the effect of having $\nu_{\bf{q}}$ either near or far from resonance with the decaying two-level atom (frequency separation $\omega$), and to check the importance of the size of the partitions via the parameter $\epsilon$.
 
Since $|\gamma_0\rangle$ is a pure state, the appropriate entanglement measure between partitions A and B of the system is the entropy of entanglement $E=S(\rho_A)$, where $S(\rho_A)=-{\rm{tr}}[\rho_A\log_2(\rho_A)]$ is the von-Neumann entropy with the reduced state $\rho_A={\rm{tr}}_B[\rho_{AB}]$. It must be emphasized that the entropy of entanglement is a entanglement monotone only if the global state is pure. Even though the pure field state $|\gamma_0\rangle$ is achieved only in the limit $t\gg \Gamma^{-1}$, we will see later on this paper that our results are still approximately valid for finite times. This broadens the applicability of our work. The reduced state for the partition $A$ can be obtained from (\ref{g0}) by tracing out modes in partition $B$. One finds
\begin{eqnarray}\label{rho}
\rho_{A}&=&\sum_{\bf{k}_j}|p_j|^2|\{0\}_A\rangle\langle _A\{0\}|+\sum_{\bf{k}_m,\bf{k}_n}p_m p_n^{*}|1_{{\bf{k}_m}},\{0\}\rangle\langle 1_{{\bf{k}_n}},\{0\}|,
\end{eqnarray} 
where $\bf{k}_j$ refers to a wave vector of some mode in the partition $B$, $\bf{k}_m$ ($\bf{k}_n$) to some mode in partition $A$, $|\{0\}_A\rangle$ to vacuum states of modes in partition $A$, $|1_{{\bf{k}_{m(n)}}},\{0\}\rangle$ means one photon in mode $\bf{k}_{m(n)}$ of partition $A$ and vacuum for the rest of the modes in that partition, and
\begin{eqnarray}
p_i=\frac{g_{\bf{k}_i}({\bf{r}}_0)}{(\nu_{k_i}-\omega)+i\Gamma/2}. 
\end{eqnarray}
The only non-zero eigenvalues of $\rho_A$ are $\lambda_1=\sum_{\bf{k}_j}|p_j|^2$ and $\lambda_2=\sum_{\bf{k}_n}|p_n|^2$, where $\bf{k}_j$ refers to partition $B$ and $\bf{k}_n$ to partition $A$. As mentioned before, the final results must be obtained by passing to the continuum. In spherical coordinates we have \cite{scully}
\begin{eqnarray}
\sum_{\bf{k}}\rightarrow 2\frac{V}{(2\pi)^3}\int_0^{2\pi}d\phi\int_0^{\pi}d\theta\sin\theta\int_0^{\infty}dk\, k^2,
\end{eqnarray}
and then
\begin{eqnarray}
\sum_{\bf{k}_j}|p_j|^2\rightarrow\frac{\wp_{ab}}{6\pi^2\hbar\epsilon_0 c^3}\left[\int_{0}^{\nu_q-\epsilon}\frac{\nu^3d\nu}{(\nu-\omega)^2+\Gamma^2/4}+\int_{\nu_q+\epsilon}^{\infty}\frac{\nu^3d\nu}{(\nu-\omega)^2+\Gamma^2/4}\right].
\end{eqnarray}
For consistency with the Weisskopf-Wigner used in the derivation of the state (\ref{g0}), we should again consider that $\nu^3$ varies little around $\nu_k=\omega$, what allows us now to replace $\nu^3$ by $\omega^3$ in the above integrals as well as to extend the lower integration limit of the first integral to $-\infty$ \cite{scully}. Making this approximations one obtains
\begin{eqnarray}
\sum_{\bf{k}_j}|p_j|^2\rightarrow 1-\frac{1}{\pi}\arctan\left[\frac{2}{\Gamma}(\epsilon+\nu_q-\omega)\right]-\frac{1}{\pi}\arctan\left[\frac{2}{\Gamma}(\epsilon-\nu_q+\omega)\right].\label{B}
\end{eqnarray}
Now, we sum the modes referring to partition $A$
\begin{eqnarray}
\sum_{\bf{k}_n}|p_n|^2\rightarrow\frac{\wp_{ab}}{6\pi^2\hbar\epsilon_0 c^3}\left[\int_{\nu_q-\epsilon}^{\nu_q+\epsilon}\frac{\nu^3d\nu}{(\nu-\omega)^2+\Gamma^2/4}\right].
\end{eqnarray}
Again, we replace $\nu^3$ by $\omega^3$ (but leave the integration limits unaltered) to obtain
\begin{eqnarray}
\sum_{\bf{k}_n}|p_n|^2\rightarrow \frac{1}{\pi}\arctan\left[\frac{2}{\Gamma}(\epsilon+\nu_q-\omega)\right]+\frac{1}{\pi}\arctan\left[\frac{2}{\Gamma}(\epsilon-\nu_q+\omega)\right].\label{A}
\end{eqnarray}
It is important to look into normalization of (\ref{g0}) because we must end up with a physical state after performing the approximations. In fact, the normalization has been conserved since the sum of (\ref{B}) with (\ref{A}) is equal to one for any values of $\epsilon$, $\omega$ and $\nu_q$. With (\ref{B}) and (\ref{A}), one can now easily obtain the entropy of entanglement $S=-\sum_{i=1}^2\lambda_i\log_2\lambda_i$ and study the bipartite entanglement between partitions $A$ and $B$. In general, two special features of the entanglement in the field modes should be highlighted, namely its dependence upon the size of the partitions and upon the detuning between the central frequency of partition $A$ and the atom $\Delta=\nu_q-\omega$. From now on, we will use the dimensionless quantities $\tilde{\epsilon}\equiv\epsilon/\Gamma$ and $\tilde{\Delta}\equiv\Delta/\Gamma$.

\begin{figure}[h]
\begin{center}
\includegraphics[scale=0.7]{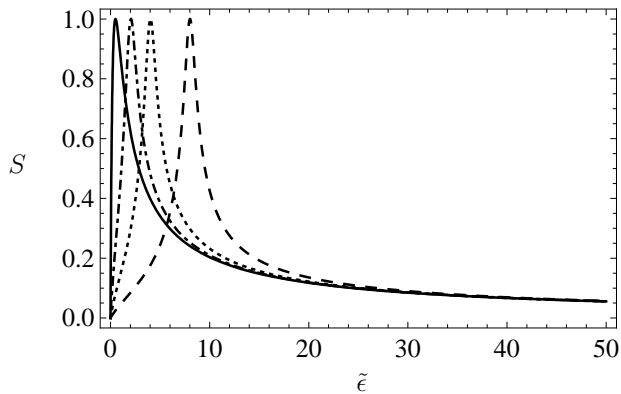}
\caption{Entanglement between partitions $A$ and $B$ as a function of the size of partition $A$. The curves corresponds to different values of $\tilde{\Delta}$: $0.0$ (solid), $2.0$ (dash-dot), $4.0$ (dot), $8.0$ (dash).}
\label{fig1}
\end{center}
\end{figure}

In Fig.(\ref{fig1}), we show how the entanglement varies with the size of partition $A$. We can see that the effect of increasing the size is initially to increase the entanglement between both partitions. It physically means that more and more entangled modes are shared by the partitions. One would expect the entanglement to saturate because as $\tilde{\epsilon}$ increases it comes to a point where the number of modes in each partition optimizes the bipartite entanglement. If $\tilde{\epsilon}$ continues to increase, the situation returns to be unbalanced with less and less entangled modes shared by the partitions. In the limit $\tilde{\epsilon}\rightarrow \infty$, there are no modes left in partition $B$, and the entanglement between the partitions naturally goes to zero. In Fig.(\ref{fig1}), one can also see that if the central frequency $\nu_{\bf{q}}$ moves from resonance with the atom (increase of $|\tilde{\Delta}|$), it takes more and more modes (larger values of $\tilde{\epsilon}$) to achieve maximum entanglement as expected. 

\begin{figure}[h]
\begin{center}
\includegraphics[scale=0.7]{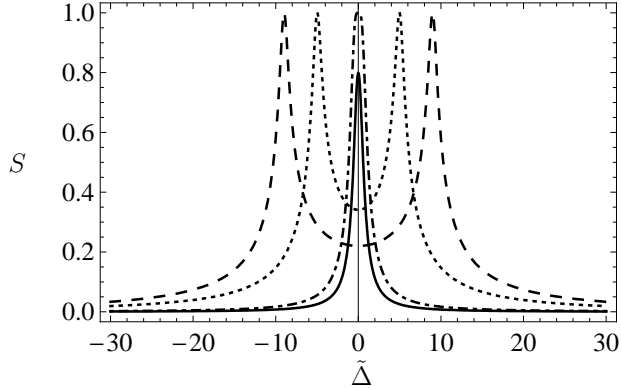}
\caption{Entanglement between partitions $A$ and $B$ as a function of the detuning between the atom and the central mode of partition $A$. The curves corresponds to different values of $\tilde{\epsilon}$: $0.2$ (solid), $0.5$ (dash-dot), $5.0$ (dot), $9.0$ (dash).}
\label{fig2}
\end{center}
\end{figure}

We have just seen that the entanglement between two partitions of the free-space modes after atomic decay considerably depends on the detuning between the atom and the central mode of partition $A$. Such dependence is show in Fig.(\ref{fig2}). First, one can notice that if the size of the frequency band $\tilde{\epsilon}$ is small, the maximum entanglement is not achieved for any detuning. This is in full accordance with all that has been already discussed here about the role played by $\tilde{\epsilon}$. On the other hand, if $\tilde{\epsilon}$ is sufficiently high, there is always a value of $\tilde{\Delta}$ which allows the system to achieve the maximum entanglement. It is physically expected that if one is to consider a partition whose central frequency is far from resonance with the atom, the modes will almost not be affected by the atomic decay. In fact, one can see in Fig.(\ref{fig2}) that in the limit $\tilde{\Delta}\rightarrow\pm\infty$ the entanglement goes to zero. We should make it clear that $\tilde{\Delta}$ can not actually be considered too big since we are in the scope of nonrelativistic quantum mechanics in which high energy interactions are not properly accounted. Indeed, even the Hamiltonian (\ref{H}) obtained in the dipole approximation would not be valid in such regime. In spite of that, the Weisskopf-Wigner theory used here is very accurate in the optical-microwave domain and fully explain the main features of atomic spontaneous emission in free space. An attempt to go beyond the Weisskopf-Wigner theory is presented in \cite{wwb}. It is also important to remark that momentum entanglement of the atom and the field is not included in our treatment. Such problem has been solved in \cite{chan,chan1} where the authors conclude that such entanglement is very small indeed.
 
The Weisskopf-Wigner theory then allows one to learn about the entanglement in the field modes after atomic spontaneous emission. Since this theory also gives the time evolution of the global state comprising the atom \textit{and} the field, one can go a step further and obtain the dynamics of entanglement between these two subsystems as well. From Eqs.(\ref{psit}), (\ref{coef}), and (\ref{ga}), we may easily obtain the reduced density operator for the atom (tracing out the modes), and it is given by
\begin{eqnarray}
\rho_{\rm{at}}(t)&=&e^{-\Gamma t}|a\rangle\langle a|\nonumber\\ &&+\sum_{\bf{k}}\left(\frac{|g_{\bf{k}}|^2+|g_{\bf{k}}|^2 e^{-\Gamma t/2}-2e^{-\Gamma t/2}|g_{\bf{k}}|^2\cos[(\nu_k-\omega)t]}{(\nu_k-\omega)^2+\Gamma^2/4}\right)|b\rangle\langle b|.
\end{eqnarray} 
Just like before, we now pass to a continuum of modes, and the result after integration is
\begin{eqnarray}
\rho_{\rm{at}}(t)=e^{-\Gamma t}|a\rangle\langle a|+(1-e^{-\Gamma t})|b\rangle\langle b|,
\end{eqnarray}
which coincides with the atomic density operator obtained in the master equation formalism in the Born-Markov approximations \cite{carmichael}. In  situations where the Markov approximation breaks, the Weisskopf-Wigner approach might be quite useful for the the study open quantum systems. An interesting problem where the Weisskopf-Wigner theory applies is inhibition of atomic spontaneous emission in photonic cristals which is known to be a non-Markovian process \cite{peter}. Another special feature of the Weisskopf-Wigner approach is that it allows us to study the state of the reservoir as well. For atomic spontaneous emission in free space, the study of the reservoir was performed in the first part of this paper.

\begin{figure}[h]
\begin{center}
\includegraphics[scale=0.7]{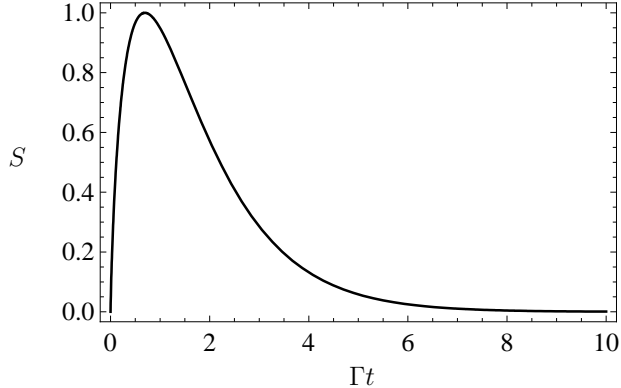}
\caption{Time evolution of the entanglement between the atom and the electromagnetic modes of free space.}
\label{fig3}
\end{center}
\end{figure}

The reduced system has now just two states, and the evaluation of the entropy of entanglement is straightforward. Its time evolution is shown in Fig.(\ref{fig3}). 
The maximum of entanglement takes place at $t_{\rm{m}}=\ln(2)/\Gamma$. Consequently, we can reinterpret the maximum of entanglement as the physical quantity that fix the time needed for half population in an atomic ensemble to decay spontaneously (this time is called \emph{half life} in nuclear physics). At $t_{\rm{m}}$, and for the particular problem treated here, the atom is in a Bell state with the reservoir $[S(\Gamma t_{\rm{m}})=1]$. 

At first sight, the results presented here seem to come into conflict with what is expected in the Born-Markovian approximations which is known to be valid for atomic spontaneous emission in free space. For example, the Born approximation states that the atom-reservoir state becomes only slightly different from a perfect separable state \cite{carmichael}. Consequently, the entanglement shown in Fig.(\ref{fig3}) should not be so high. In fact, if we were to evaluate the entanglement between the atom and each mode or with a small group of modes, we would see that there is really just a small deviation from product states. However, when we consider the whole bunch of free space modes, this bipartite entanglement may become surprisingly high. Consequently, even though it is right to say that the atom-reservoir state is almost separable due to the fact the reservoir is a very large system (not much affected by the atom), it is exactly because it is that big that the small entanglement between each mode and the atom leads to high entanglement when all modes are considered, as shown in Fig.(\ref{fig3}). We think this a central result in our paper. The validity of the factorization assumption is quantitatively studied in a perturbative approach in \cite{cunha}.

\begin{figure}[h]
\begin{center}
\includegraphics[scale=0.7]{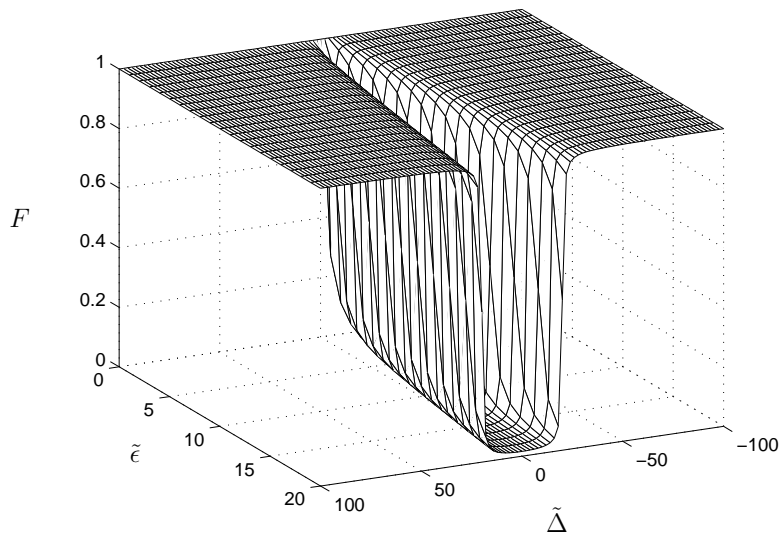}
\caption{Fidelity between the state of partition $A$ and the vacuum as an function of the size of the partition and the detuning between its central mode and the atomic frequency.}
\label{fig4}
\end{center}
\end{figure}

When we think of Markov approximation in this problem, i.e. the independence of the future behavior of $\rho_{\rm{at}}(t)$ on its past history, it is usual to justify it on the following reasonable physical grounds \cite{carmichael}. If the reservoir is a large quantum system, it is expected that it is mantained in thermal equilibrium, so that it is not supposed to preserve the changes induced by the atom at previous times. Since we are dealing with the zero temperature case, the thermal equilibrium state reads the vacuum state. Now, it is our findings in the first part of this paper, i.e. the study of the entanglement properties of the field state after spontaneous emission, that seem to come into conflict with the Markov approximation. Again, the size of the reservoir is key for understanding it. The field state after atomic decay (\ref{g0}) contains just one photon \emph{shared} by an infinity of radiation modes. We physically expect that this state is in fact very close to the vacuum state at the beginning of the atomic decay process (except for the fact that it is an entangled state). If this is true, the Markov approximation is still valid because of the very small deviation from the thermal equilibrium. A naive calculation of the fidelity $F={\rm{Tr}(|\gamma_0\rangle\langle \gamma_0|\rho_{\rm{vacuum}}})$, would lead to $F=0$. This is not relevant for the kind of problem we are dealing with (infinitely many subsystems). For example, we expect the states $|0,1\rangle$ to be ``more different" from $|0,0\rangle$ than $|0,0,0,0,0,0,0,0,0,0,0\rangle$ is from $|0,0,0,0,0,0,0,1,0,0,0\rangle$. Basically, our expectation is based on the number of subsystems in the vacuum state. The more subsystems in the vacuum, the more such a state is close to the global vacuum state. However, the fidelity is zero for both cases in this example. Again, it is physically more appealing to concentrate our attention in one partition instead of the global state $|\gamma_0\rangle$. In Fig.(\ref{fig4}), we show how the state (\ref{rho}) of the partition $A$  compares to the vacuum state by evaluating  $F(\tilde{\epsilon},\tilde{\Delta})\equiv{\rm{Tr_A}(\rho_A\rho_{\rm{vacuum}}}$) in the continuum limit. We can see that there is a vast region in which the fidelity is equal to one with respect to the vacuum. Although it is not a formal proof that the field state stays at equilibrium, one can easily see that depending on the \emph{cut} or partition, the resulting state is in fact very close to the vacuum. 

We would like to return to the discussion of the validity of our results for finite times. The entanglement properties of the field state studied in this paper were obtained under the assumption that the overall state of the modes was pure. We now argue that this high purity is still approximately true even for finite times. We first recall that the distillable entanglement $E_D$ \cite{distillable} is lower bounded by entropies as follows \cite{martin}
\begin{eqnarray}
E_D(\rho_{AB})\leq\max[S(\rho_{B})-S(\rho_{AB}),0],
\end{eqnarray}
where, in our case, $A$ and $B$ are partitions of the field modes. In the special instance of a overall pure field state $[S(\rho_{AB})=0]$, the above inequality becomes an equality and entanglement is quantified by the entropy of the reduced state, as we did in this work. In order for our results to be valid (at least qualitatively), we must make sure that the entropy of the field modes $S(\rho_{AB})=0$ is in fact approximately zero. From the Araki-Lieb inequality \cite{book} $S(\rho_{{\rm{at}},AB})\le |S(\rho_{AB})-S(\rho_{{\rm{at}}})|$, and the fact that the atom-field state (\ref{psit}) remains pure all the time, we see that $S(\rho_{AB})=S(\rho_{{\rm{at}}})$.   Fig.(\ref{fig3}) clearly shows that entropy of the atom (consequently the entropy of the modes $AB$) goes to zero quite rapidly. For example, $S(\rho_{AB})=10^{-2}$ is achieved for $\Gamma t\approx 7$. As one can see, our results remain essentially valid even for finite times, and this fact may facilitates the experimental investigation of the entanglement properties discussed here.

To summarize, we have studied bipartite entanglement in the atomic spontaneous emission phenomenon. For the field state after atomic decay, we have found that very detuned modes weakly entangle with the rest, and that the size of the frequency band of each partition strongly changes the entanglement. We have carefully analyzed the behavior of entanglement when such parameters change, and also physically justified all features. It is important to remark that the present work represents a study about fundamentals of matter-field interactions from the point of view of quantum information, and it does not necessarily aim at applications. Actually, applications of the entanglement in the free space radiation field state after simple atomic decay as considered here seems unlikely. Of course, atomic decay in free space could be used to entangle distant atoms as shown in \cite{dist,dist1}. However, what we have just studied here is entanglement in the free space modes. This makes coherent control a challenging task when compared to atomic entanglement. 

On the other hand, we believe that a scheme to experimentally access the field entanglement induced by atomic decay could make use of photonic crystals. Spontaneous emission in this environment has been studied in \cite{peter}, and since this system is much more controllable than the electromagnetic fields in free space, we feel that this could be an alternative to study photonic entanglement due to atomic decay. It is hoped that the present work may be a starting point for addressing this question (theoretical treatment of entanglement properties of atomic decay in photonic crystals) as well as other interesting related questions such as the application of the techniques used in \cite{wwb} to study atomic decay. 
\section*{Acknowledgements}
We would like to thank the referees for providing valuable suggestions aimed at the improvement of our manuscript. This work was performed as part of the Brazilian
National Institutes of Science and Technology (Instituto Nacional de Ci\^encia e Tecnologia - Informa\c c\~ao Qu\^antica, INCT-IQ). This work is partially supported by CNPq (Conselho Nacional de Desenvolvimento Cient\'\i fico e Tecnol\'ogico) (F. L. S.) and INCT (J. F. L.) fellowships. 


\begin{thebibliography}{00}
\bibitem{dense}C. H. Bennett and S. J. Wiesner, Phys. Rev. Lett. {\bf 69}, 2881 (1992).
\bibitem{tele} C. H. Bennett, G. Brassard, C. Cr\'epeau, R. Jozsa, A. Peres, and W. K. Wootters, Phys. Rev. Lett. {\bf 70} 1895 (1993)
\bibitem{martin} M. B. Plenio and S. Virmani, Quant. Inf. Comp. \textbf{7}, 1 (2007). 
\bibitem{qpt} L. Amico, R. Fazio, A. Osterloh, and V. Vedral, Rev. Mod. Phys. \textbf{80}, 517 (2008).
\bibitem{qpt2} T. R. de Oliveira, G. Rigolin, M. C. de Oliveira, and E. Miranda, Phys. Rev. Lett. {\bf 97}, 170401 (2006). 
\bibitem{popescu}S. Popescu, A. J. Short, and A. Winter, Nature Phys. \textbf{2}, 754 (2006).
\bibitem{lloyd}S. Lloyd, Science \textbf{319}, 1209 (2008).
\bibitem{eisert} J. Eisert, M. Cramer, M.B. Plenio, e-print quant-ph/0808.3773. 
\bibitem{ww}V. Weisskopf and E. Wigner, Z. Phys. \textbf{63}, 54 (1930). 
\bibitem{scully}M. O. Scully and M. S. Zubairy, \textit{Quantum Optics} (Cambridge University Press, United Kingdom, 1997). 
\bibitem{petrosyan}P. Lambropoulos and D. Petrosyan, \textit{Fundamentals of Quantum Optics and Quantum Information} (Springer-Verlag, Berlin, 2007).
\bibitem{harmonic} K. Audenaert, J. Eisert, M. B. Plenio, and R. F. Werner, Phys. Rev. A \textbf{66}, 042327 (2002).
\bibitem{kg} M. B. Plenio, J. Eisert, J. Drei\ss ig, and M. Cramer, Phys. Rev. Lett. \textbf{94}, 060503 (2005).
\bibitem{wolf} N. Schuch, J. I. Cirac, and M. M. Wolf, Commun. Math. Phys. \textbf{267}, 65 (2006). 
\bibitem{unanyan} R. G. Unanyan, M. Fleischhauer, and D. Bru\ss, Phys. Rev. A \textbf{75}, 040302 (2007).
\bibitem{gioev} D. Gioev and I. Klich, Phys. Rev. Lett. \textbf{96}, 100503 (2006).
\bibitem{barthel}T. Barthel, M.-C. Chung, and U. Schollw\"{o}ck, Phys. Rev. A \textbf{74}, 022329 (2006).
\bibitem{w} X. -B. Chen, N. Zhang, S. Lin, Q. -Y Wen, F. -C. Zhu, Opt. Commun. \textbf{281}, 2331 (2008).
\bibitem{wwb} S. R. Zhao, C. P. Sun, and W. X. Zhang, Phys. Lett. A \textbf{207}, 327 (1995).
\bibitem{chan} K.W. Chan, C. K. Law, and J. H. Eberly, Phys. Rev. Lett. \textbf{88}, 100402 (2002).
\bibitem{chan1} K. W. Chan, C. K. Law, and J. H. Eberly, Phys. Rev. A \textbf{68}, 022110 (2003).
\bibitem{carmichael}H. Carmichael, \textit{An Open systems Approach to Quantum Optics} (Springer-Verlag, Berlin, 1993).
\bibitem{peter} D. G. Angelakis, P. L. Knight, and E. Paspalakis, Contemp. Phys. \textbf{45}, 303 (2004).
\bibitem{cunha} M. O. Terra Cunha, S. Geraij Mokarzel, J. G. Peixoto de Faria, and M. C. Nemes, e-print quant-ph/0410049. 
\bibitem{distillable} C. H. Bennett, D. P. DiVincenzo, J. A. Smolin, and W. K. Wootters, Phys. Rev. A \textbf{54}, 3824 (1996).
\bibitem{book} G. Jaeger, \textit{Quantum Information: An Overview } (Springer, New York, 2007).
\bibitem{dist} S. Bose, P.L. Knight, M.B. Plenio and V. Vedral, Phys. Rev. Lett. 83, 5158 (1999).
\bibitem{dist1} C. Cabrillo, J. I. Cirac, P. Garcia-Fernandez, and P. Zoller, Phys. Rev. A {\bf{59}}, 1025 (1999).
\end{thebibliography}
\end{document}